\documentclass{PoS}

\newcommand{\beqn}{\begin{eqnarray}}
\newcommand{\eeqn}{\end{eqnarray}}
\newcommand{\eq}[1]{(\ref{#1})}
\newcommand{\M}{PLSM${}_q$}

\newcommand{\dd}{{\mathrm d}}

\newcommand{\Tr}{{\mathrm{Tr}\,}}
\newcommand{\Z}{{\mathbb{Z}}}
\newcommand{\Dirac}{\rlap {\hspace{-0.5mm} \slash} D}

\title{Possible splitting of deconfinement and chiral transitions in strong magnetic fields in QCD}

\ShortTitle{Possible splitting of deconfinement and chiral transitions}

\author{\speaker{Eduardo S. Fraga}\\
Instituto de F\'isica, Universidade Federal do Rio de Janeiro,
Caixa Postal 68528, Rio de Janeiro, RJ 21941-972, Brazil\\
E-mail: \email{fraga@if.ufrj.br}}

\author{Ana J\'ulia Mizher\\
Centro Brasileiro de Pesquisas Fisicas, Rua Xavier Sigaud 150, 22290-180 Rio de Janeiro, RJ, Brazil\\
E-mail: \email{anajulia@cbpf.br}}

\author{M.~N.~Chernodub\thanks{On leave from ITEP, Moscow, Russia.}\\
CNRS, Laboratoire de Math\'ematiques et Physique Th\'eorique,
Universit\'e Fran\c{c}ois-Rabelais, F\'ed\'eration Denis Poisson - CNRS,
Parc de Grandmont, Universit\'e de Tours, 37200 France\\
Department of Physics and Astronomy, University of Gent, Krijgslaan 281, S9, B-9000 Gent, Belgium\\
E-mail: \email{maxim.chernodub@lmpt.univ-tours.fr}}

\abstract{We show that finite-temperature deconfinement and chiral transitions can split in
a strong enough magnetic field. The splitting in critical temperatures of these transitions
in a constant magnetic field of a typical LHC magnitude is of the order of 10 MeV.
A new deconfined phase with broken chiral symmetry appears.}

\FullConference{35th International Conference of High Energy Physics - ICHEP2010,\\
        July 22-28, 2010\\
        Paris France}

\begin{document}

The thermal quark-hadron transition can be dramatically modified in the presence of a {\it strong}
magnetic field \cite{Agasian:2008tb,Fraga:2008qn,Boomsma:2009yk,Fukushima:2010fe,ref:main,Gatto:2010qs}.
We work in the two-flavor linear sigma model coupled to quarks and
to the Polyakov loop (\M) in the presence of an external magnetic field \cite{ref:main}.
The confining properties of QCD are described by the  complex-valued
Polyakov loop variable $L$. The expectation value of the Polyakov
loop $L$ is an {\it exact} order parameter of the color confinement
in the limit  of infinitely massive quarks:
\beqn
\mbox{Confinement}:\quad
\left\{
\begin{array}{llll}
\langle L \rangle  & = & 0 \quad , \quad & \mbox{low $T$}   \\
\langle L \rangle  & \neq & 0 \quad , \quad & \mbox{high $T$}
\end{array}
\right.\,, \qquad\quad L(x) = \frac{1}{3} \Tr {\cal P} \exp \Bigl[i
\int\limits_0^{1/T} \dd \tau \, A_4(\vec x, \tau) \Bigr]\,,
\label{eq:L} \label{eq:L:phases} \eeqn
where $A_4 = i A_0$ is the matrix-valued temporal component of the Euclidean
gauge field $A_\mu$ and the symbol ${\cal P}$ denotes path ordering. The
integration takes place over compactified imaginary time $\tau$.

The chiral features of the model are encoded in the dynamics of the
$O(4)$ chiral field, the singlet component of which is an exact order parameter in the chiral limit
(i.e., when quarks and pions are massless degrees of freedom):
\beqn
\mbox{Chiral symmetry}:\quad
\left\{
\begin{array}{llll}
\langle \sigma \rangle  & \neq & 0 \quad , \quad & \mbox{low $T$}   \\
\langle \sigma \rangle  & = & 0 \quad ,\quad & \mbox{high $T$}
\end{array}
\right.\,,
\label{eq:sigma:phases}
\qquad\qquad
\begin{array}{lll}
\phi & = & (\sigma,\vec{\pi})\,,\\
\vec{\pi} & = & (\pi^{+},\pi^{0},\pi^{-})\,.
\end{array}
\label{eq:phi}
\eeqn
Here $\vec{\pi}$ is the isotriplet of the pseudoscalar pion fields and
$\sigma$ is the chiral scalar field which plays the role of an approximate
order parameter of the chiral transition in QCD.

The quark field $\psi$ provides interaction between the Polyakov loop $L$ and the chiral field
$\phi$, making a bridge between confining and chiral properties.
Quarks are also coupled to the external magnetic field
since the $u$ and $d$ quarks are electrically charged. Thus, it is clear
that the external magnetic field will affect the chiral dynamics {\it and}
the confining properties of the model.

The Lagrangian of \M\  describes the constituent quarks $\psi$,
which interact with the meson fields $\sigma$, $\pi^\pm = (\pi^1 \pm i \pi^2)/\sqrt{2}$ and $\pi^0 = \pi^3$,
the Abelian gauge field $a_\mu = (a^0,\vec a) = (0, - B y,0,0)$,
and the $SU(3)$ gauge field $A_\mu$
(related to the Polyakov loop $L$, see Eq.~\eq{eq:L}) via the covariant derivative
$\Dirac = \gamma^{\mu} (\partial _{\mu} - i Q\, a_\mu - i A_\mu)$ with the charge matrix
$Q = {\mathrm{diag}} (+ 2e/3, -e/3)$:
\beqn
{\cal L} =  \overline{\psi} \left[i \Dirac - g(\sigma +i\gamma _{5}
 \vec{\tau} \cdot \vec{\pi} )\right]\psi + \frac{1}{2}\bigl[(\partial _{\mu}\sigma)^2 + (\partial _{\mu} \pi^0)^2\bigr]
 + |D^{(\pi)}_\mu|^2 - V_\phi(\sigma ,\vec{\pi}) - V_L(L,T)\,,
\label{eq:L:full}
\eeqn
where $D_\mu^{(\pi)} = \partial_\mu + i e a_\mu$ is the covariant derivative acting on colorless pions.
The chiral potential is
\beqn
V_\phi(\sigma ,\vec{\pi}) = \frac{\lambda}{4}(\sigma^{2}+\vec{\pi}^{2} - {\it v}^2)^2-h\sigma\,,
\qquad
h = f_{\pi} m_{\pi}^2\,, \quad v^2 = f^2_\pi-{m^{2}_{\pi}}/{\lambda}\,, \quad \lambda = 20\,,
\eeqn
where $f_\pi \approx 93\,\mbox{MeV}$ and $m_\pi \approx 138\,\mbox{MeV}$.
The constituent quark mass is given by
$m_q \equiv m_q(\langle\sigma\rangle)= g \langle\sigma\rangle$,
and, choosing $g=3.3$ at $T=0$, one obtains for the constituent quarks in the
vacuum $m_q \approx 310~$MeV. At low temperatures quarks are not
excited, and the model reproduces results from the usual linear
$\sigma$-model without quarks.

The Polyakov potential is given by \cite{ref:ratti08}
\beqn
\frac{V_L(L,T)}{T^4} =-\frac{L^*L}{2}\sum_{l=0}^2 a_l \left(\frac{T_0}{T}\right)^l + b_3\left(\frac{T_0}{T} \right)^3\,\ln\left[1-6\,L^*L+4\left({L^*}^3+L^3\right) - 3\left(L^*L\right)^2\right]\,,
\label{eq:VL}
\eeqn
where $T_0 \equiv T_{SU(3)} = 270\, \mbox{MeV}$ is the critical temperature
in the pure gauge case and $a_0 = 16\,\pi^2/45 \approx 3.51$, $a_1 = -2.47$, $a_2 = 15.2$, and $b_3 = -1.75$.
Below we follow a mean-field analysis in which the mesonic sector is treated classically whereas quarks represent fast degrees of freedom.

\begin{figure}[!thb]
\begin{center}
\begin{tabular}{cc}
\includegraphics[width=73mm,clip=true]{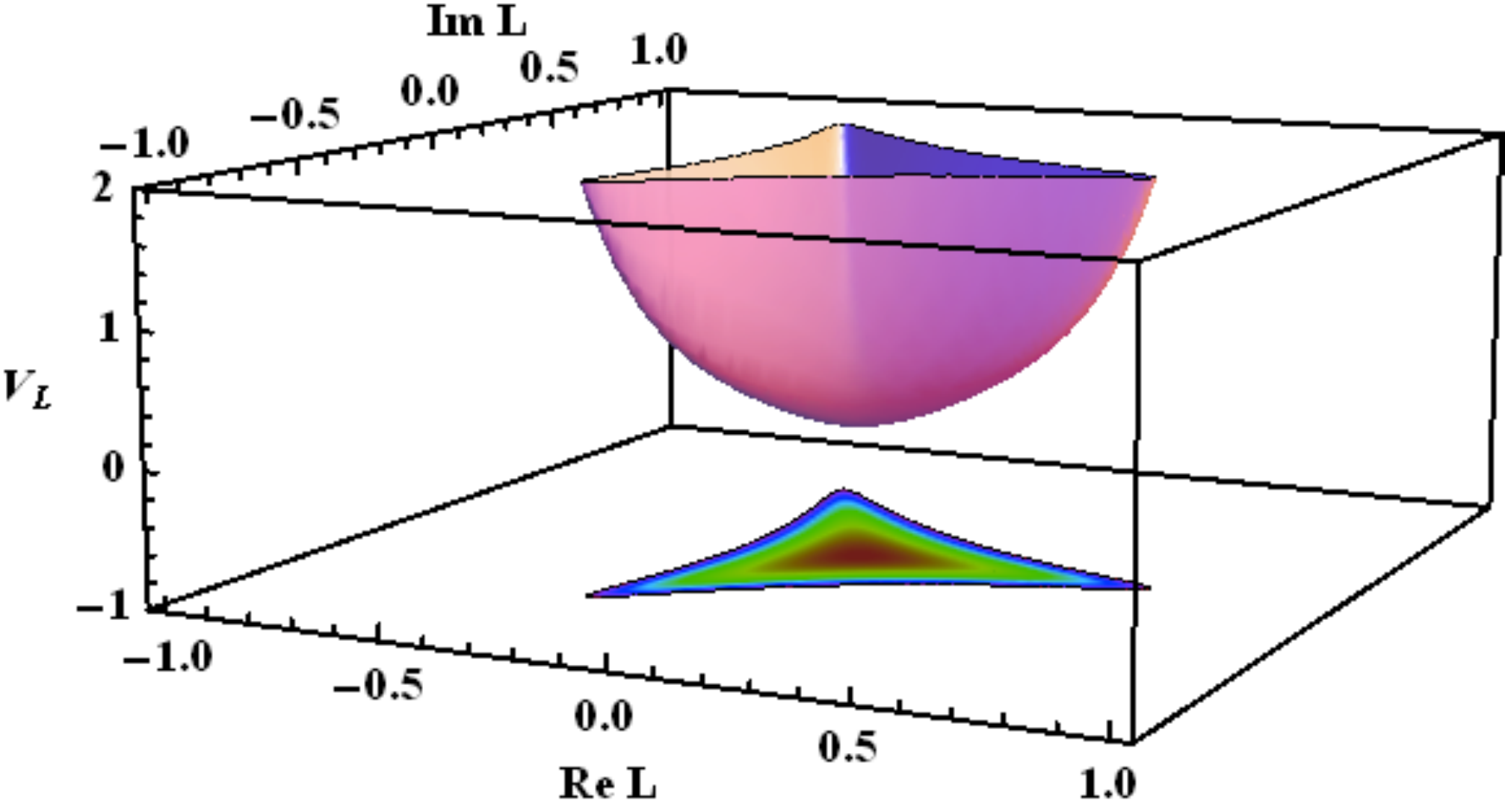} &
\includegraphics[width=75mm,clip=true]{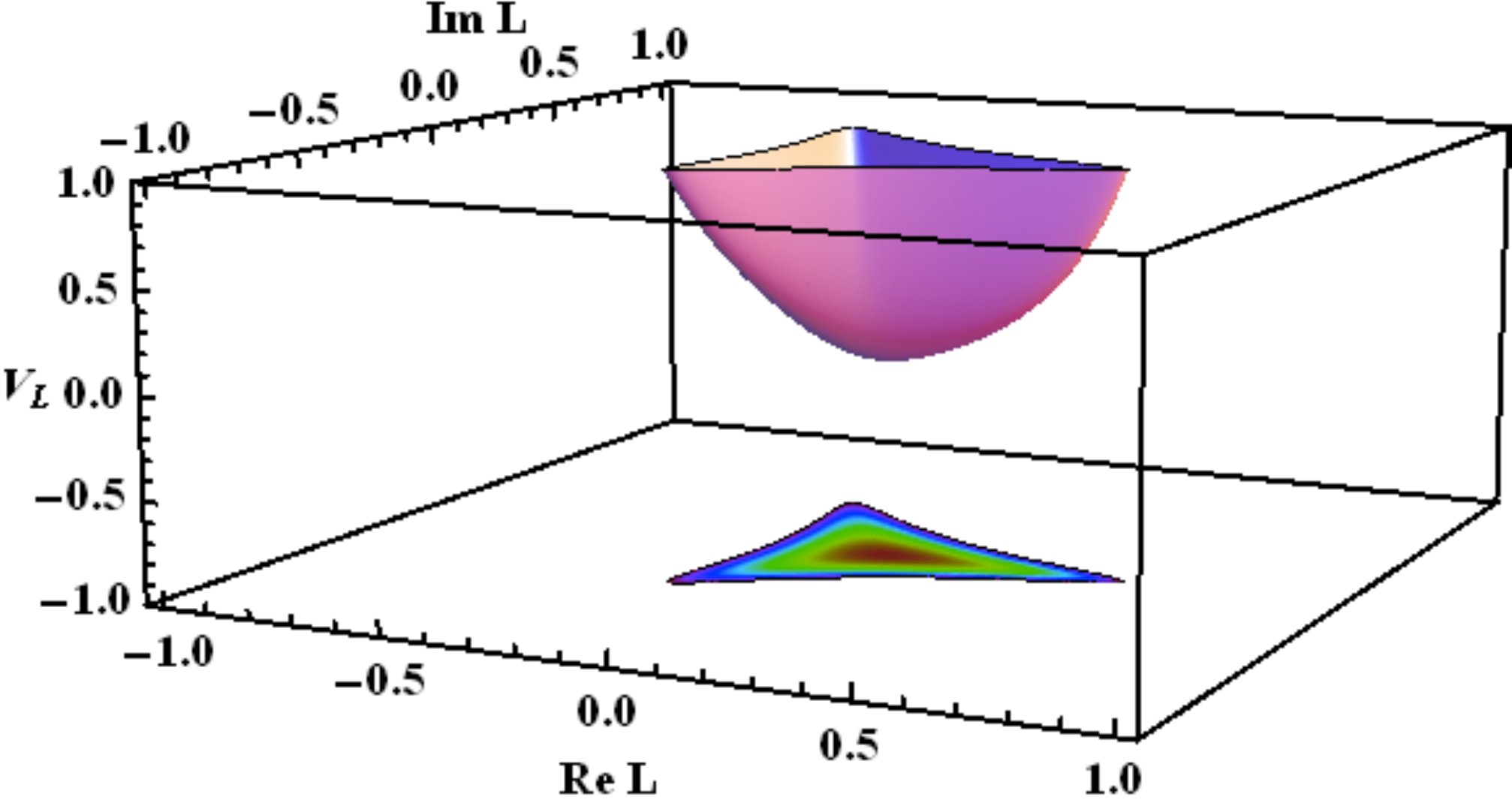} \\[-43mm]
\hskip 35mm {\large $T=0.8T_{c}$,\quad  $eB= 0$} &
\hskip 35mm {\large $T=0.8T_{c}$, $eB= 9 T^2$} \\[43mm]
\includegraphics[width=73mm,clip=true]{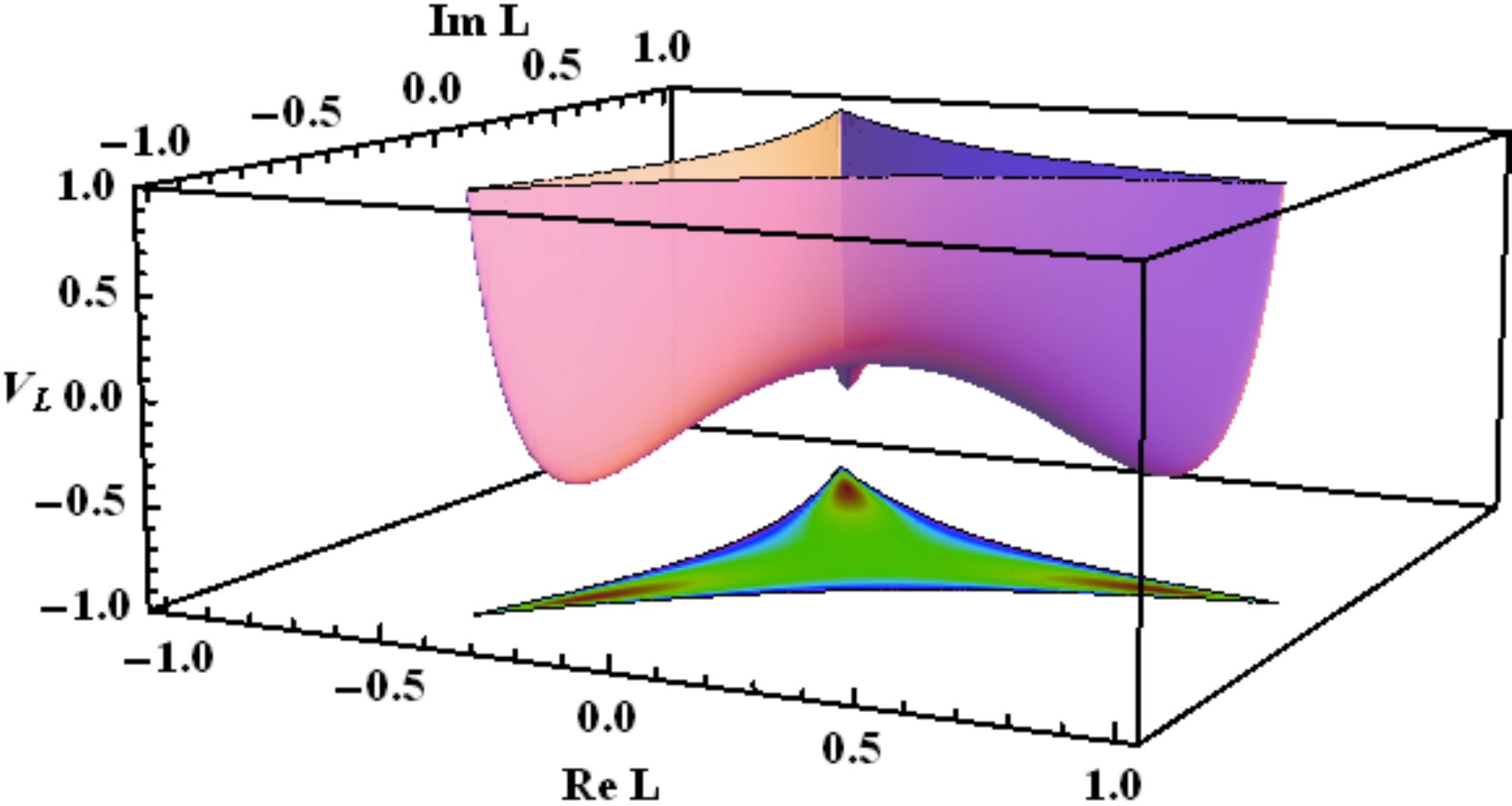} &
\includegraphics[width=75mm,clip=true]{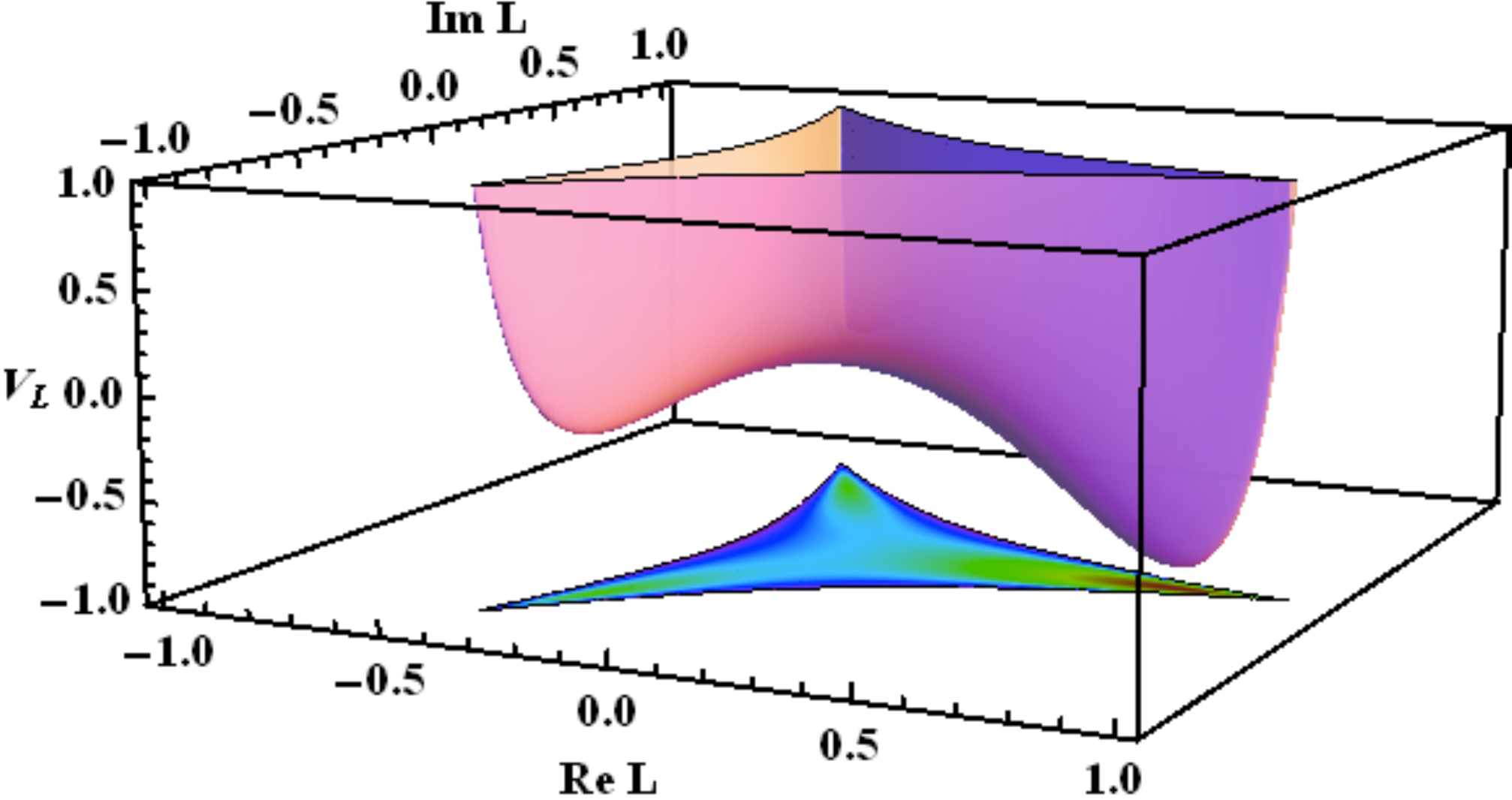} \\[-43mm]
\hskip 35mm {\large $T=1.2T_{c}$, $eB = 0$} &
\hskip 35mm {\large $T=1.2T_{c}$, $eB= 9 T^2$}\\[35mm]
\end{tabular}
\end{center}
\caption{Effects of temperature and magnetic field on quark confinement:
The Polyakov loop potential at $T=0.8 \, T_0$ (top) and $T=1.2 \, T_0$ (bottom) and at zero magnetic field (left) and at $e B = 9 T^2$ (right).}
\label{fig:potential}
\end{figure}

In the confining sector, the strong magnetic field affects the potential for the expectation value
of the Polyakov loop via the intermediation of the quarks in three ways \cite{ref:main}:
\begin{itemize}
\item[(i)] the presence of the magnetic field intensifies the breaking of the global $\Z_3$ symmetry and
makes the Polyakov loop real-valued (see Fig. 1);
\item[(ii)] the thermal contribution from quarks tends to destroy the confinement phase by increasing the
expectation value of the Polyakov loop;
\item[(iii)] on the contrary, the vacuum quark contribution tends to restore the confining phase by
lowering the expectation value of the Polyakov loop.
\end{itemize}

\begin{figure}[!thb]
\begin{center}
\begin{tabular}{cc}
\qquad no vacuum corrections & \qquad with vacuum corrections \\
\includegraphics[width=70mm,clip=true]{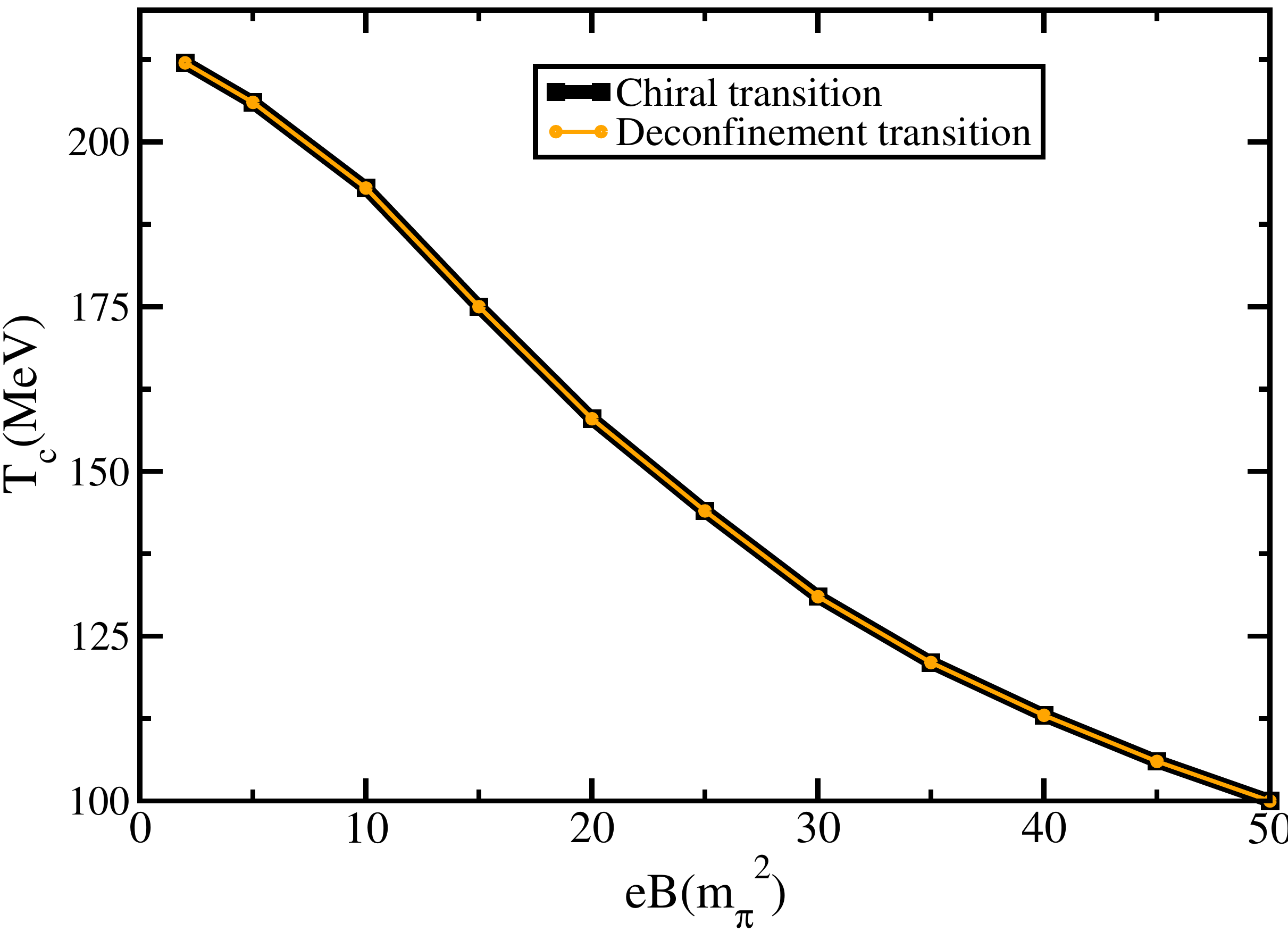} 
& \includegraphics[width=70mm,clip=true]{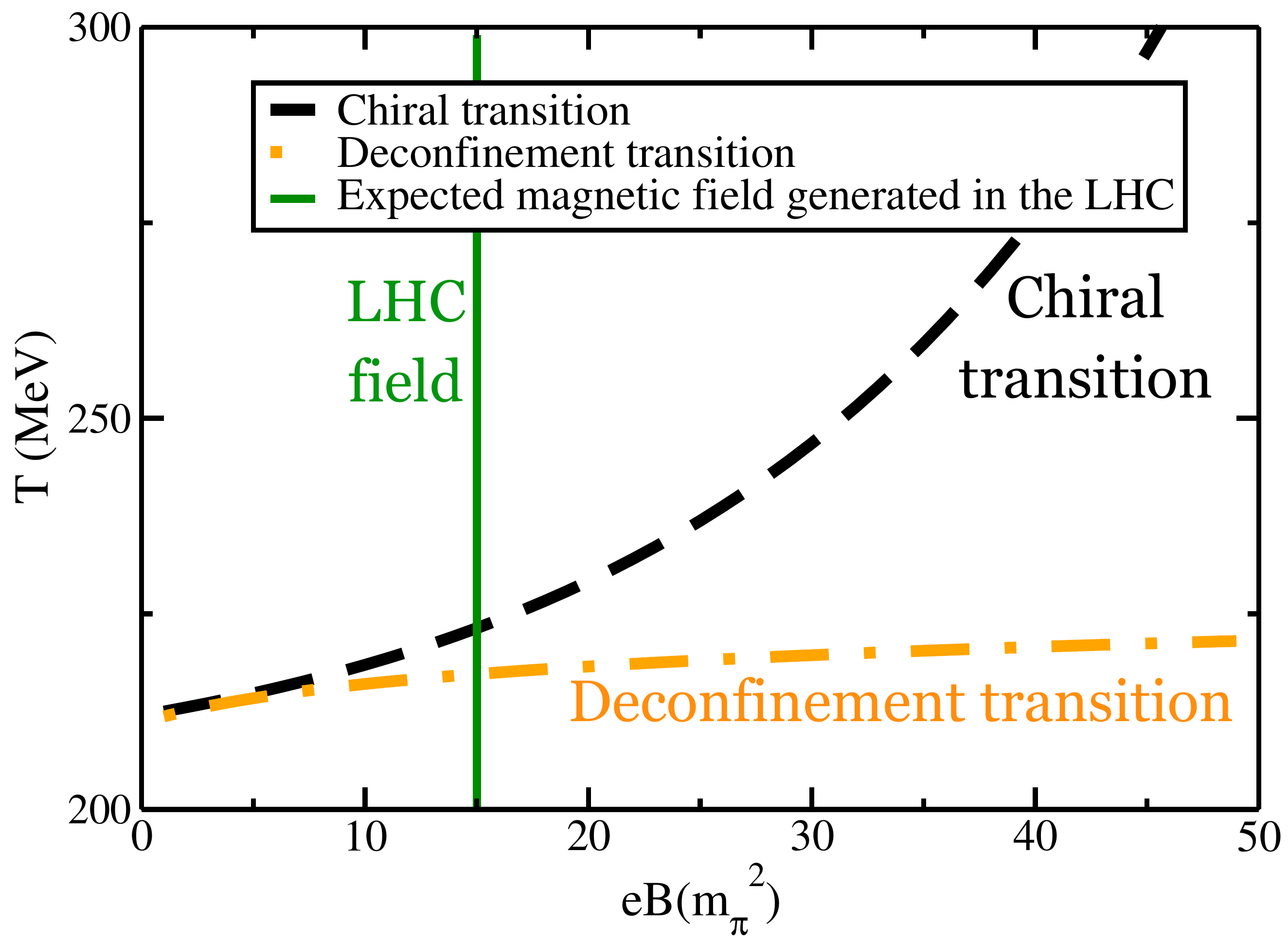}\\[-3mm]
\end{tabular}
\end{center}
\caption{Phase diagram in the $B$-$T$ plane.
(left) Without vacuum corrections: the critical temperatures of the deconfinement (the dash-dotted line)
and chiral (the dashed line) transitions coincide all the way, and decrease with $B$.
(right) With vacuum corrections:  the critical temperatures of the deconfinement (the dash-dotted line)
and chiral (the dashed line) coincide at $B=0$ and split at higher values of the magnetic field. A deconfined 
phase with broken chiral symmetry appears. The vertical line represents a typical
magnitude of the magnetic field that expected to be realized at LHC heavy-ion collisions~\cite{Skokov:2009qp}.}
\label{fig:phase:diagram}
\end{figure}

The vacuum correction from quarks has a crucial impact on the phase structure. If one ignores
the vacuum contribution from the quarks, then one finds that the confinement and chiral phase
transition lines coincide, and in this case the increasing magnetic field lowers the common
chiral-confinement transition temperature \cite{ref:main}. However, if one includes the vacuum
contribution, then the confinement and chiral transition lines split, and both chiral and deconfining
critical temperatures become increasing functions of the magnetic field, Fig.~\ref{fig:phase:diagram}.
The vacuum contribution from the quarks affect drastically the chiral sector as well.
Our calculations also show that the vacuum contribution seems to soften the order of the phase
transition: the first-order phase transition -- which would be realized in the absence of the vacuum
contribution -- becomes a smooth crossover in the system with vacuum quark loops included.

Either scenario is very exciting, and brings new possibilities for the phase diagram of strong
interactions: a first-order order transition, the splitting of coexistence lines, new phases, the
explicit breaking of the center $\Z_3$ symmetry by the magnetic field, and so on. 

An independent investigation
in the Nambu-Jona-Lasinio model agrees with the second scenario \cite{Gatto:2010qs}, 
which was expected since by definition this model includes quark degrees of freedom 
in the vacuum. Preliminary numerical simulations of lattice QCD~\cite{D'Elia:2010nq} 
seems to favor the second scenario as well 
indicating that the critical temperatures of both transitions are rising with strength of the background magnetic field.

\acknowledgments
This work was partially supported by CAPES-COFECUB, CNPq, FAPERJ, and FUJB/UFRJ.

\end{document}